\documentclass[%
 reprint,
 amsmath,amssymb,
 aps,
]{revtex4-2}

\usepackage{subcaption}

\usepackage{graphicx}
\usepackage{dcolumn}
\usepackage{bm}

\begin{document}

\preprint{APS/123-QED}

\title{Implementation and Deployment of an Injection Tuning Tool \\Using Bayesian Optimization at the SuperKEKB Accelerator}

\author{Shinnosuke Kato}
 \email{s-kato@hep.phys.s.u-tokyo.ac.jp}
 \affiliation{The University of Tokyo, Bunkyo, Tokyo 113-0033, Japan}

\author{Gaku Mitsuka}
 \affiliation{KEK, Oho, Tsukuba, Ibaraki 305-0801, Japan,}
 \affiliation{SOKENDAI, Shonan Village, Hayama, Kanagawa 240-0193, Japan}

\date{\today}

\begin{abstract}

As of July 2025, the SuperKEKB accelerator, which collides 7 GeV electrons with 4 GeV positrons to abundantly produce particles such as B mesons and $\tau$ leptons, holds the world record for the highest instantaneous luminosity. Continuous operation and upgrades are underway to achieve even higher luminosities. Maintaining a high instantaneous luminosity requires sustaining high beam currents in the storage rings, which in turn demands efficient beam injection from the injector. In particular, a high injection efficiency, defined as the ratio of the beam current successfully accumulated in the ring to the current delivered from the beam transport line, must be ensured. In the present study, we developed a tool to automate the injection tuning process using Bayesian optimization, a machine-learning-based technique, in order to improve the injection efficiency. During test operations conducted in November–December 2024, this tool successfully enhanced the injection efficiency by up to 32\%.

\end{abstract}

\maketitle


\section{Introduction}

\begin{figure}[htbp]
\centering
\includegraphics[width=\linewidth]{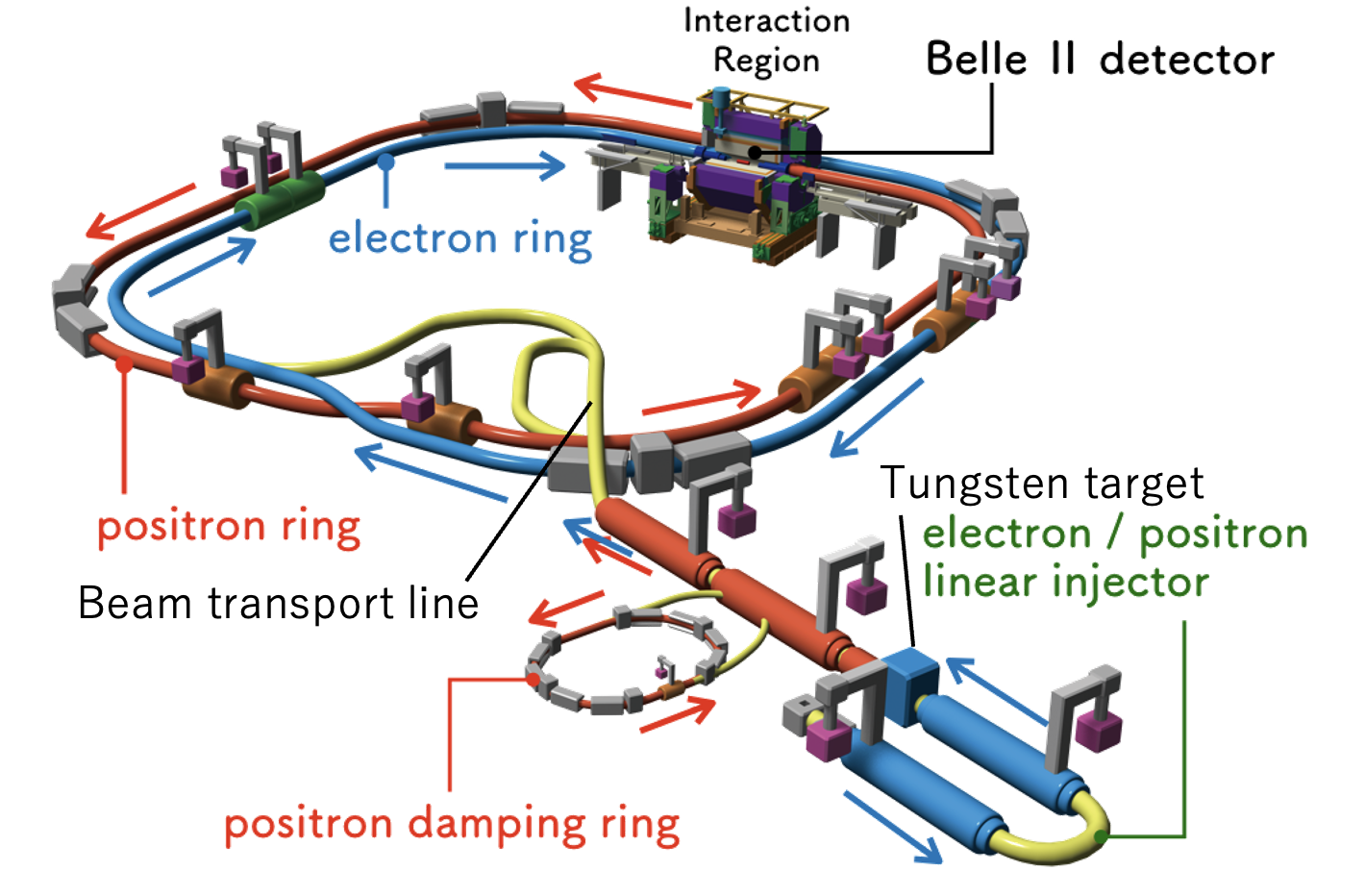}
\caption{Overview of the SuperKEKB accelerator~\cite{SuperKEKB:TDR_overview}}
\label{fig:SuperKEKB_fig_english}
\end{figure}

The SuperKEKB accelerator is an electron–positron collider that, as of July 2025, achieved a world-record instantaneous luminosity of $5.1\times10^{34}~\text{cm}^{-2}\text{s}^{-1}$. It consists of two separate storage rings, each with a 3~km circumference: a high-energy ring (HER), which stores 7~GeV electrons, and a low-energy ring (LER), which stores 4~GeV positrons. The ring structure comprises four repeated sections, each containing straight and arc sections. Particles are continuously injected into the main ring and collide at the Belle II detector. An overview of the SuperKEKB accelerator is shown in Fig.\ref{fig:SuperKEKB_fig_english}.

Achieving further records of instantaneous luminosity at SuperKEKB requires advanced optimization of the beam-injection system. Conventionally, injection tuning has been performed manually by operators, who rely on their personal knowledge and expertise to adjust a wide range of parameters. However, as the tuning process becomes increasingly complex, continued reliance on operator-driven adjustments raises concerns about the growing time and effort required to maintain stable operation. To address this issue, this study introduces machine-learning techniques to develop an automated injection-tuning tool that assists human operators in the optimization process.

As of July 2025, no simulation has accurately reproduced the real behavior of the injection system at SuperKEKB, rendering the injection-tuning process an effective black-box function. Consequently, it is difficult to predict in advance which set of parameters will yield the optimal injection efficiency without directly measuring the performance. Moreover, the optimal parameter set is subject to unpredictable time variations on the order of a few hours, and previously effective settings have often failed to deliver comparable performances when reused. In addition, the measurement uncertainty of the injection efficiency is relatively large, typically 2–2.5\%, which has been reported to cause the classical black-box optimization algorithm, the Downhill Simplex method~\cite{10.1093/comjnl/7.4.308}, to fail frequently~\cite{NatsuiLCWS2024}. Furthermore, large parameter changes may cause the beam to deviate from stable circulation conditions, potentially leading to beam aborts.

To address the aforementioned challenges associated with manual injection tuning, this study developed an automated injection-tuning tool that utilizes machine-learning techniques. This is the first attempt to incorporate machine learning into injection tuning during physics runs at a collider. The tool developed in this study is designed to meet the following two performance requirements.

\begin{itemize}
\item \textbf{Improvement of injection efficiency}: The tool should be capable of improving the injection efficiency in real time using only the current injection parameter settings. Even in cases where optimization does not lead to further improvement, the existing performance must not be degraded.
\item \textbf{Ensuring operational safety}: The tool must be designed to avoid triggering beam aborts, thereby maintaining a high level of operational safety.
\end{itemize}

The remainder of this paper is organized as follows. 
Section~\ref{Injection Scheme into the Main Ring} provides an overview of the injection scheme for the main ring.
Section~\ref{Bayesian Optimization} describes the Bayesian optimization method employed in this study, along with additional newly implemented features for its application in injection tuning.
Section~\ref{Optimization Results} presents the results of various optimization studies using Bayesian optimization, as well as the outcomes of the test operations of the injection-tuning tool.
Finally, Section~\ref{Conclusion} summarizes the conclusions of this study.

\section{Injection Scheme into the Main Ring}
\label{Injection Scheme into the Main Ring}

\begin{figure}[htbp]
\centering
\includegraphics[width=\linewidth]{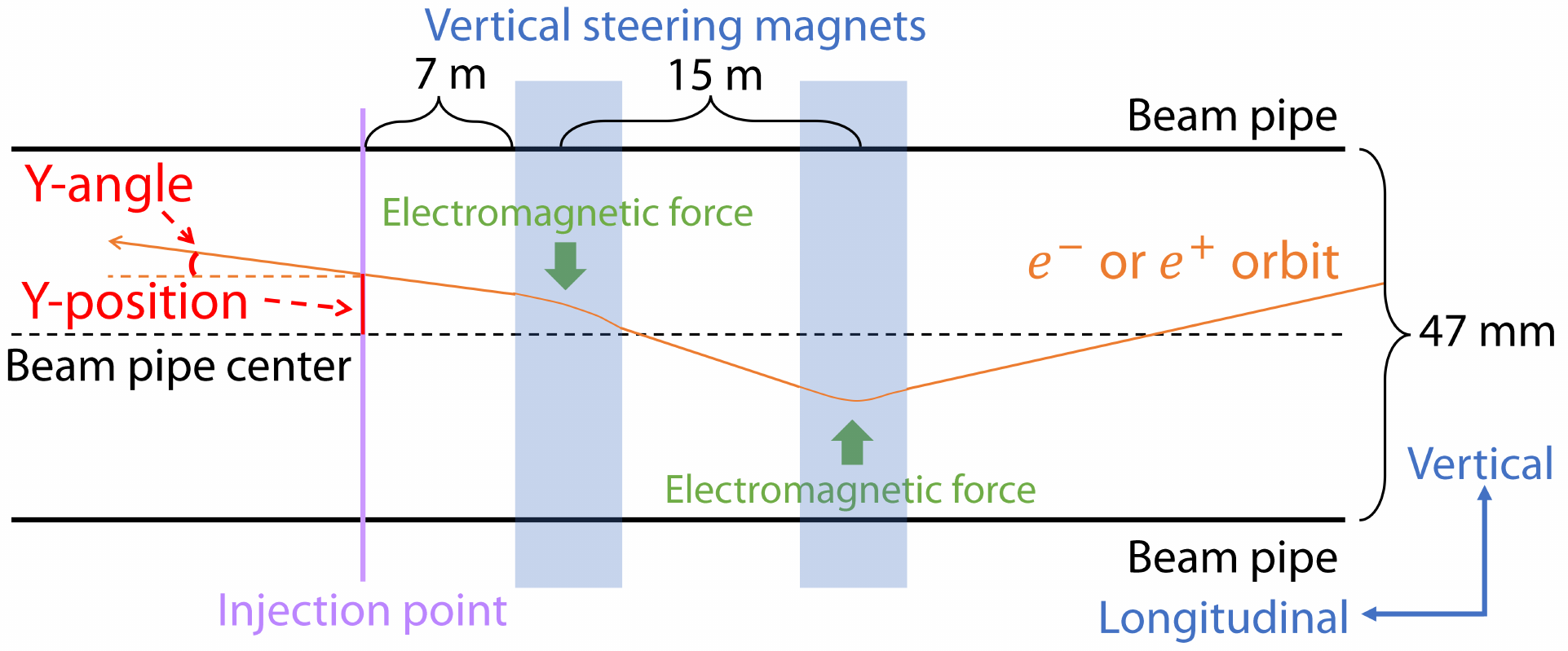}
\caption{Schematic of the injection system in the vertical plane.}
\label{fig:Vstexplain_english}
\end{figure}

\begin{figure}[htbp]
\centering
\includegraphics[width=\linewidth]{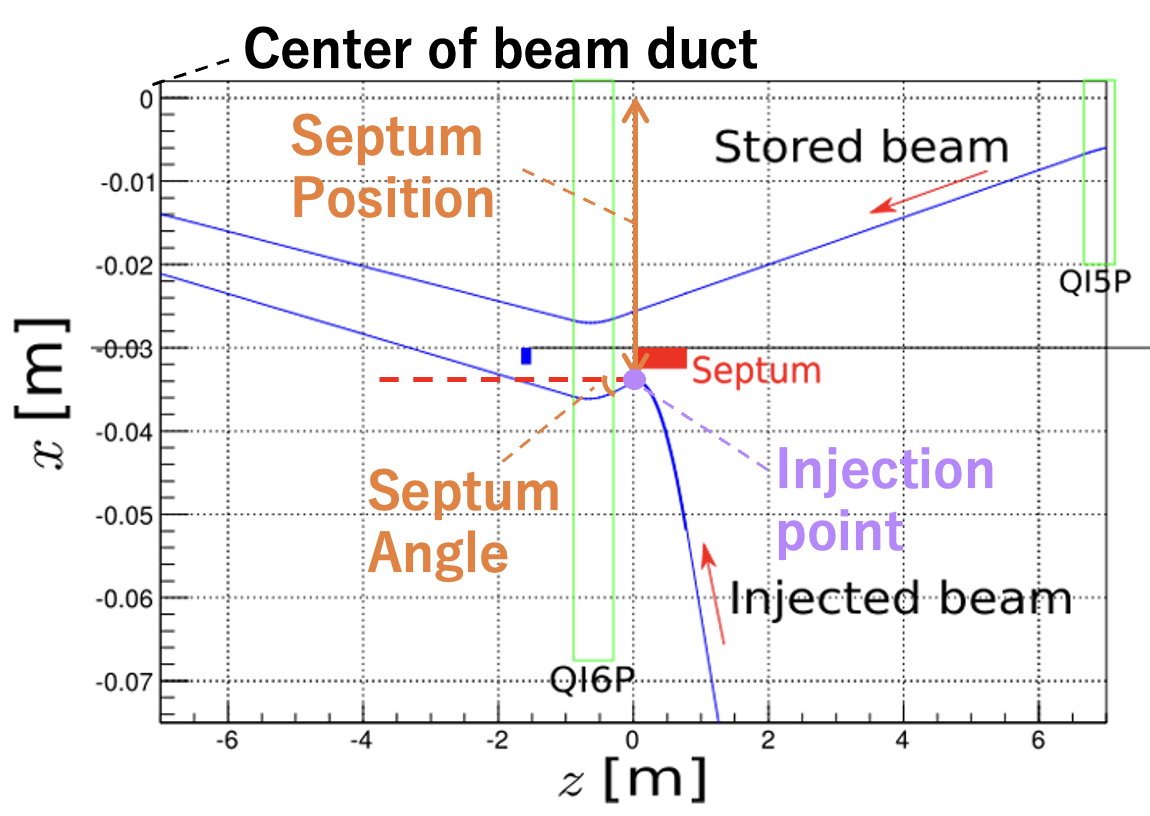}
\caption{Schematic of the injection system in the horizontal plane~\cite{SuperKEKB:TDR_BT}.}
\label{fig:Septumexplain_new}
\end{figure}

This section describes the injection-tuning method from the downstream section of the beam transport line to the main ring.
Figures~\ref{fig:Vstexplain_english} and \ref{fig:Septumexplain_new} show the injection systems in the vertical and horizontal planes, respectively, with respect to the beam direction.

\begin{figure}[htbp]
    \centering
    \begin{subfigure}[b]{0.48\linewidth}
        \centering
        \includegraphics[width=\linewidth]{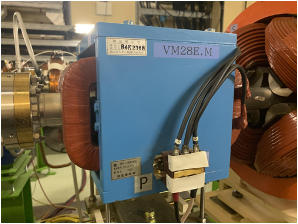}
        \caption{Vertical steering magnet}
        \label{fig:vst}
    \end{subfigure}
    \hfill
    \begin{subfigure}[b]{0.48\linewidth}
        \centering
        \includegraphics[width=\linewidth]{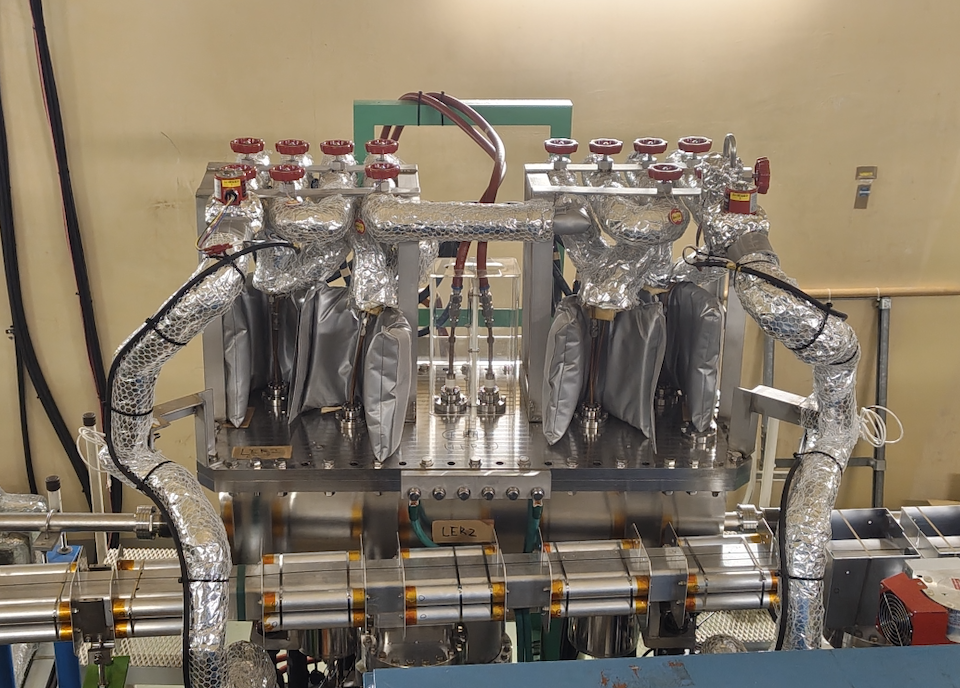}
        \caption{Septum magnet}
        \label{fig:septum}
    \end{subfigure}
    \caption{Photograph of the magnets used for injection tuning in this study.}
    \label{fig:gaikan_inj}
\end{figure}

The injected bunch arriving from the beam transport line is first adjusted in the vertical plane using a vertical steering magnet to control its vertical position and angle at the injection point. Subsequently, a septum magnet is used to adjust the horizontal position and angle relative to the center of the beam duct at the same location. A quadrupole magnet is then used to bend the orbit toward the center of the beam duct in the horizontal plane. Finally, a kicker magnet is employed to match the orbits of both the injected and stored bunches to the circulating orbit. A photograph of the magnets used in this study is shown in Fig.~\ref{fig:gaikan_inj}.

The injection-tuning parameters used in this study are as follows. Because the injection schemes for the LER and HER are identical, the discussion in this section focuses on the LER as a representative example.

As illustrated in Fig.~\ref{fig:Vstexplain_english}, two pairs of vertical steering magnets are located downstream of the beam transport line. The vertical position and angle of the injected beam could be controlled by adjusting the excitation currents of these magnets. The vertical distance from the center of the beam duct to the injection point, located approximately 7~m downstream of the vertical steering magnets, is defined as the y-position, and the vertical angle at that point is defined as the y-angle.

As shown in Fig.~\ref{fig:Septumexplain_new}, the orbit of the injected bunch is adjusted using two parameters: Septum Position and Septum Angle. These correspond to the horizontal distance and angle of the injected bunch relative to the center of the beam duct at the injection point, which is located at the downstream end of the septum magnet. The injected bunch is bent in the direction opposite to the center of the beam duct to better match its horizontal angle with that of the stored bunch. It should be noted that the horizontal and vertical scales in Fig.~\ref{fig:Septumexplain_new} are different, because the figure is intended to show the overall geometry of the injection section. The actual deflection angle of the injected bunch by the septum magnet is on the order of 100~mrad.Although horizontal injection is primarily controlled by both septum and kicker magnets, in this paper we focus on the septum magnet.

The injection efficiency $E$ is defined by Eq.~\eqref{Eff_English}.

\begin{equation}
E = \frac{I_{\text{stored bunch, after injection}} - I_{\text{stored bunch, before injection}}}{I_{\text{injected bunch}}}
\label{Eff_English}
\end{equation}

The denominator of Eq.\eqref{Eff_English}, representing the current injected into the main ring, is measured using a stripline beam position monitor (BPM) installed along the beam transport line~\cite{striplineBPM, SuperKEKB:TDR_monitor}.
In contrast, the numerator in Eq.\eqref{Eff_English}, which corresponds to increase in the stored bunch current in the main ring, is measured using a bunch current monitor (BCM)~\cite{SuperKEKB:TDR_monitor}.

\section{Bayesian Optimization}
\label{Bayesian Optimization}

This section provides an overview of Bayesian optimization~\cite{Mockus1978, Mockus1989}, the machine-learning method employed in this study, and introduces the additional functionalities required to adapt it for injection tuning.

Bayesian optimization is a type of black-box optimization used to optimize functions whose analytic form is unknown. Assuming that the explicit form of the function is inaccessible, optimization is efficiently performed through sequential observations. This method combines Gaussian process regression~\cite{rasmussen2006gaussian}, which constructs a probabilistic model of the function, with an acquisition function that selects the next evaluation point by balancing exploitation and exploration. In combination, these components enable accurate probability-based prediction of the function’s maximum using only a limited number of observations. In the present setup, the computation time required for Bayesian optimization was negligible compared with the time required for the objective evaluation.

One of the advantages of this method is that it does not require prior training data because the model is updated in real time based on measurement data. This makes Bayesian optimization particularly suitable for online experimental optimization. Bayesian optimization can also be warm-started using prior data, and this functionality is implemented in our framework, although it was not used in the present study.

In a prior study aimed at maximizing positron yield at the KEK Linac~\cite{PhysRevAccelBeams.27.084601}, Bayesian optimization outperformed two other black-box optimization algorithms, Tree-structured Parzen Estimator (TPE)~\cite{NIPS2011_86e8f7ab} and Covariance Matrix Adaptation Evolution Strategy (CMA-ES)~\cite{hansen2023cmaevolutionstrategytutorial}, demonstrating its high sample efficiency among the three methods. Therefore, Bayesian optimization was adopted for injection tuning in this study.

To apply Bayesian optimization to injection tuning, it is essential to account for beam aborts. If the injection-tuning parameters are freely varied within a wide search space, some regions may produce increased injection-induced beam background, potentially leading to beam aborts upon entering such regions. Therefore, it is necessary to ensure that the injection-induced beam background remains below a certain threshold so that Bayesian optimization can be conducted under safe conditions.

Herein, three key features were implemented to enable the application of Bayesian optimization for the injection tuning. First, segmented region optimization was introduced as a strategy to avoid regions with high beam loss by gradually expanding the parameter domain in directions that improve the injection efficiency. Second, a step-by-step variation was introduced to prevent beam aborts, as abrupt parameter changes might lead to unsafe conditions. Third, a proximal biasing method~\cite{proximal,Roussel:2023yin} was implemented to reduce the overall optimization time by favoring exploration near previously tested parameters.

\subsection{Segmented Region Optimization}

\begin{figure}[htbp]
\centering
\includegraphics[width=\linewidth]{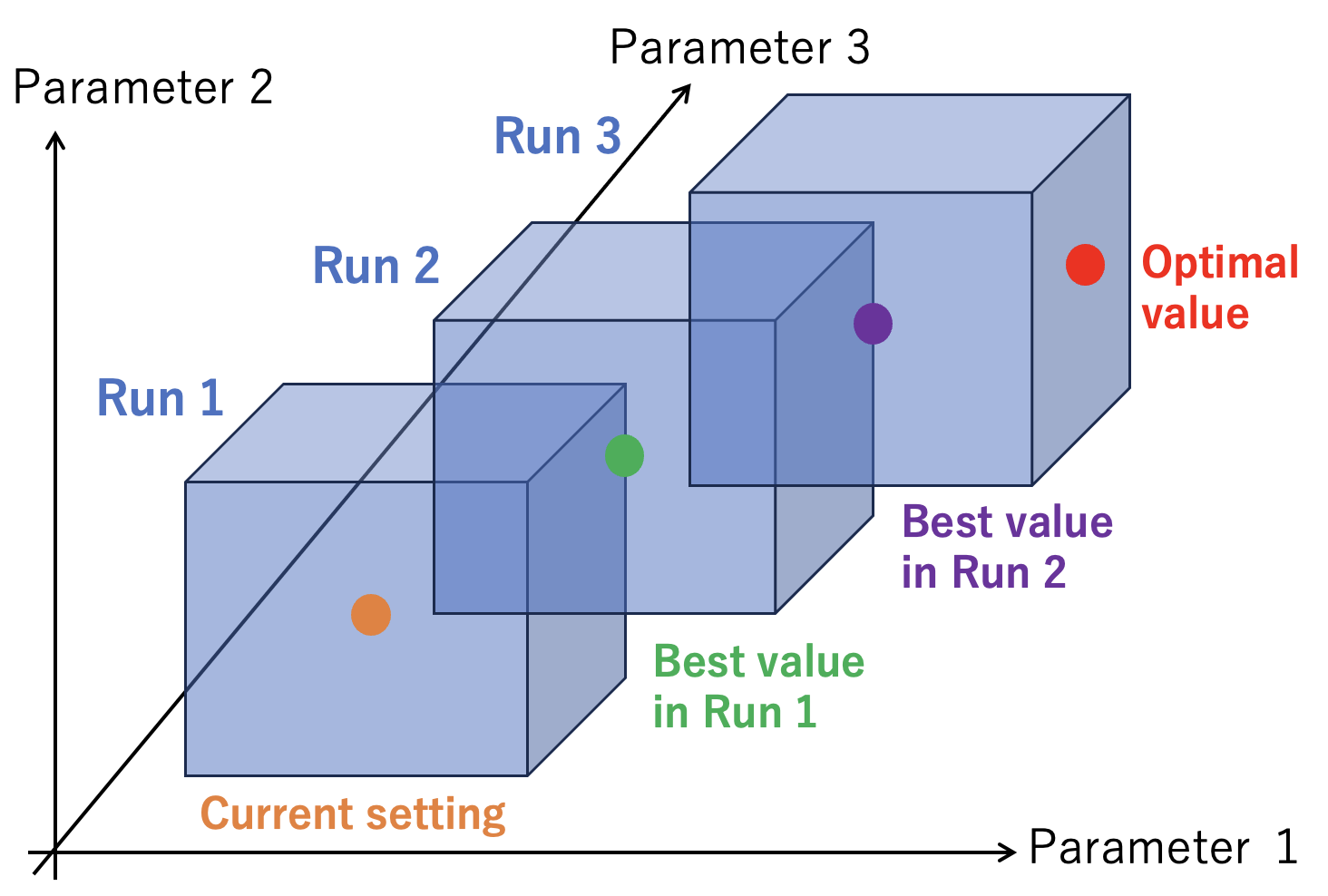}
\caption{Conceptual illustration of the segmented region optimization. The blue regions represent the parameter domains for each run, which are redefined around the parameter set that achieved the highest injection efficiency in the previous run.}
\label{fig:Kowake_english}
\end{figure}

The segmented region optimization method, illustrated in Fig.~\ref{fig:Kowake_english}, involves repeated optimizations within small subregions of the parameter space. Empirically, it is known that under stable beam conditions, no sudden beam loss occurs within a sufficiently small region centered on the current parameter set. Therefore, a small domain was initially defined around the pre-tuning parameter set, and Bayesian optimization was performed within that range. Compared with TuRBO~\cite{eriksson2019scalable}, which adaptively expands or contracts the trust region depending on the optimization performance, segmented region optimization is designed as a more conservative method. It moves the search region while keeping it within a predefined safe parameter range, rather than expanding the region aggressively. This conservative design is important because exploring unsafe parameter regions may cause beam loss or machine damage.As an initialization condition, several random points were generated within ±10\% of the parameter range centered on the pre-tuning values, and these served as the initial points for Bayesian optimization.

In this study, Bayesian optimization was performed using 10 initial sampling points generated prior to starting the optimization. A single evaluation in the optimization process was referred to as a "trial," and 30 trials were conducted for each optimization. The combination of 10 initial points and 30 optimization trials was defined as a "run." After each run, a new parameter domain was defined around the parameter set that yielded the highest injection efficiency, and the same optimization procedure was repeated for one to three additional runs. This approach enables efficient exploration over a broader parameter space by incrementally moving the search domain.

However, this method has a potential limitation. The narrowly defined parameter domain may increase the possibility of converging to a local maximum, which can potentially degrade optimization performance. Herein, the width of the parameter domain for each run was determined based on empirical knowledge obtained from manual assessments performed by experienced operators.

\subsection{Step-by-Step Variation}

\begin{figure}[htbp]
\centering
\includegraphics[width=\linewidth]{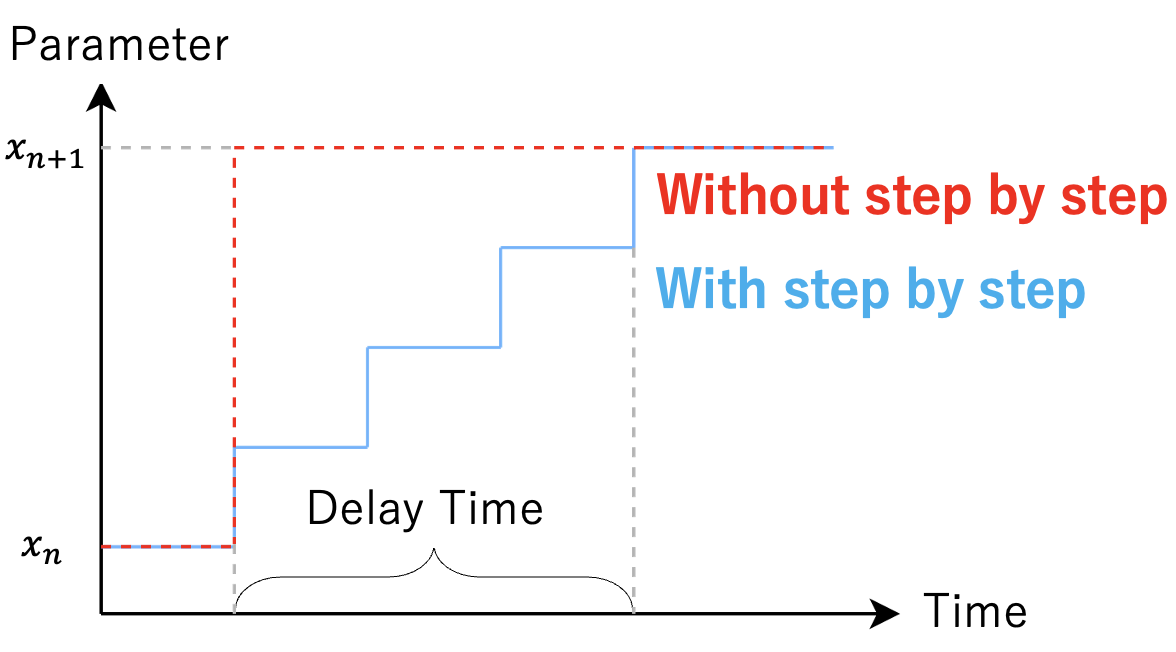}
\caption{Conceptual illustration of the step-by-step variation.}
\label{fig:StepByStep_english}
\end{figure}

The purpose of the step-by-step variation was to gradually vary the injection-tuning parameters over time, thereby avoiding large changes in the magnet settings in a single update, as illustrated in Fig.~\ref{fig:StepByStep_english}. The red dashed line represents parameter variation without the step-by-step variation, where abrupt changes occur. By contrast, the solid blue line shows how the step-by-step variation decomposes a large parameter change into smaller updates. This allows the power supply, especially the septum-magnet power supply, to reach the target setting with the required precision at each step. The step size used in the step-by-step variation was set to 0.0015 mm per step for y-position, and 0.0015 mrad per step for y-angle and Septum Angle. The number of steps was determined by dividing the total parameter change by these step-size limits. Thus, larger parameter changes require more intermediate steps, which increases the wall-clock time spent on step-by-step parameter transitions.

\subsection{Proximal Biasing Method}

\begin{figure}[htbp]
\centering
\includegraphics[width=0.9\linewidth]{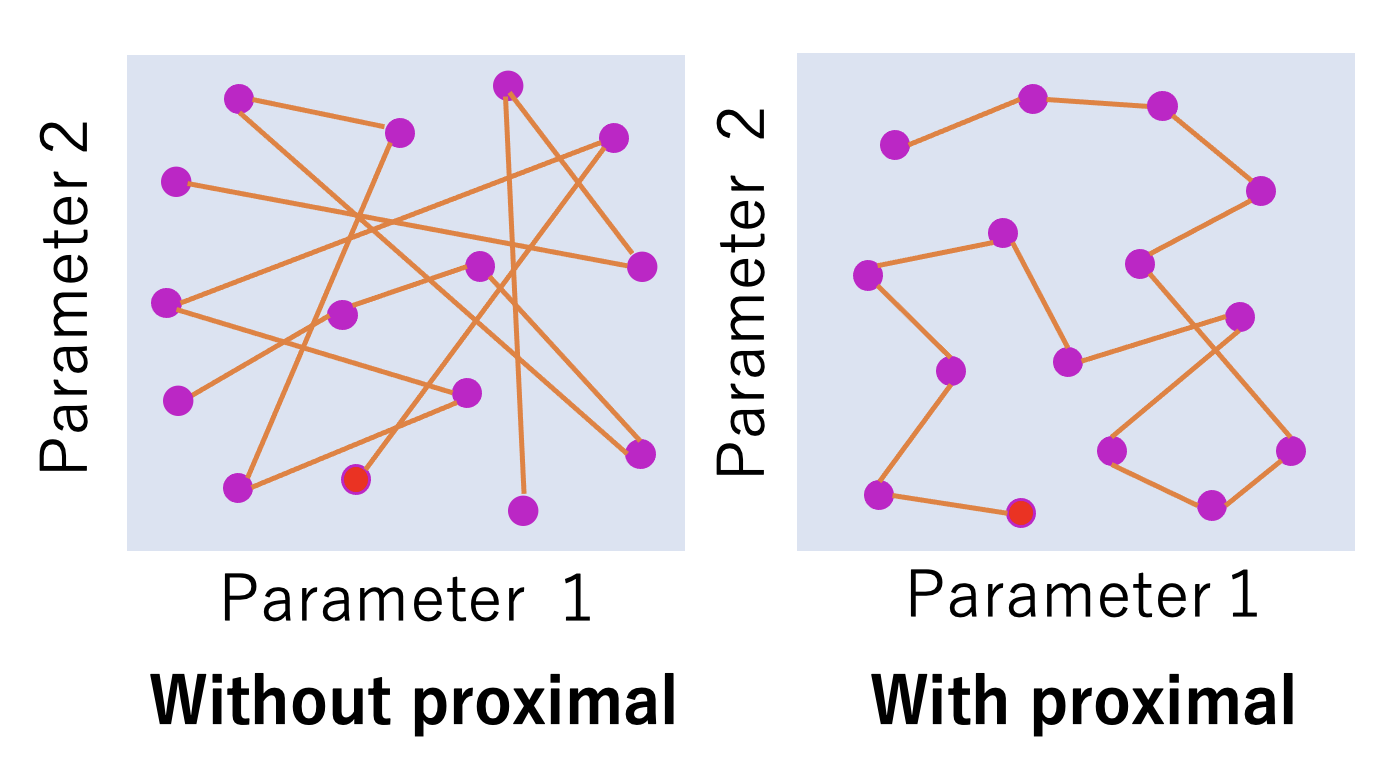}
\caption{Conceptual illustration of parameter trajectories with and without the proximal biasing method in a two-dimensional parameter space.}
\label{fig:One-stroke_english}
\end{figure}

The proximal biasing method is a technique where distance-based weighting is applied to the acquisition function, giving preference to candidate points closer to the current parameter set. This method was introduced to improve the practical efficiency of the optimization under the step-by-step operation by suppressing unnecessary long-distance parameter transitions. Figure~\ref{fig:One-stroke_english} illustrates the concept. The purple points represent candidate parameter sets selected by Bayesian optimization, and the orange lines represent the parameter transitions between successive candidate points. Each orange line consists of multiple step-by-step updates. In the left panel, large parameter transitions require many step-by-step updates and therefore longer wall-clock time for parameter movements. In contrast, in the right panel, shorter transitions require fewer step-by-step updates and reduce the wall-clock time spent on parameter movements.

Equation~\eqref{proximal_equ} represents the weighting used in the proximal biasing method

\begin{equation}
\tilde{\alpha}(\mathbf{x}^*) =
\alpha(\mathbf{x}^*)
\exp\left[
-\frac{1}{2}
\sum_{i=1}^{d}
w_i^2
\left(x_i^* - x_{0,i}\right)^2
\right]
\label{proximal_equ}
\end{equation}

Here, $\tilde{\alpha}(\mathbf{x}^*)$ denotes the acquisition function after applying the weighting,
$\alpha(\mathbf{x}^*)$ is the original acquisition function before weighting,
and $\mathbf{x}_0$ is the current parameter set. 
The farther the candidate point $\mathbf{x}^*$ is from $\mathbf{x}_0$,
the more suppressed $\tilde{\alpha}(\mathbf{x}^*)$ becomes. When $\alpha(\mathbf{x}^*)$ can take negative values, it is transformed into a non-negative value using the Softplus function.
$d$ is the number of optimization parameters, and
$\mathbf{w}=(w_1,\ldots,w_d)$ is a vector of hyperparameters that controls the degree of weighting and is referred to as the proximity bias factor. A larger proximity bias factor gives a higher priority to candidate points closer to the current parameter set. One drawback of the proximal biasing method is its potential to degrade optimization performance. In Eq.~\eqref{proximal_equ}, distance-based weighting favors candidate points near the current parameter set, which can cause the acquisition function to ignore the point that appears most promising according to the original acquisition criterion. Thus, the proximity bias factor controls the trade-off between optimization performance, such as sample efficiency or final injection-efficiency performance, and the wall-clock time spent on step-by-step parameter movements.

In principle, each component $w_i$ can be assigned independently for each optimization parameter.
In the present first implementation, however, we used the same value for all parameters.
This choice was made because the step-by-step variation already limits the rate of parameter changes,
and the same step-by-step change rate was applied to all parameters in the normalized parameter space.
Therefore, the same effective length scale was used for all dimensions in this study.

\section{Optimization Results}
\label{Optimization Results}

In this section, we present the results of injection tuning using the Bayesian optimization tool, along with several related studies. We begin with a discussion in Section~\ref{Uncertainty in Injection Efficiency} of the measurement uncertainty and statistical fluctuations associated with the injection efficiency, which serves as the evaluation metric for the optimization. Section~\ref{Verification of Reproducibility} evaluates the reproducibility of the proposed method through multiple optimization runs under identical beam conditions and parameter domains over a period of approximately 2–3~h, to examine whether the injection efficiency improves consistently. We also analyzed the parameter importance to identify the parameters that contributed to the improvement in the injection efficiency. Section~\ref{Effect of Proximal Biasing Level} presents a comparative study of different proximal biasing factors. Finally, Section~\ref{Test Deployment of the Injection Tuning Tool} summarizes the results of the test deployment of the injection-tuning tool conducted between November and December 2024.

Because this study represents the first application of machine learning to injection tuning into the main ring, we used only three parameters—y-position, y-angle, and Septum Angle—as the initial targets for optimization. These parameters were chosen for their importance and relatively low risk of equipment damage.

In this study, one Bayesian-optimization iteration typically took approximately 40--60 seconds, including the parameter transition, injection-efficiency measurement, and Bayesian-optimization computation. One optimization run took approximately 27--40 minutes.

\subsection{Uncertainty in Injection Efficiency}
\label{Uncertainty in Injection Efficiency}

This section discusses the measurement uncertainty and statistical fluctuations associated with the injection efficiency, which serve as the evaluation metrics in Bayesian optimization. In this study, the injection efficiency was measured 30 times, and the median of these 30 measurements was used as the representative value. The median was adopted to reduce the effect of occasional outliers, such as extremely low injection efficiency values, which can significantly bias the result if the mean is used. 

\begin{figure}[htbp]
\centering
\begin{minipage}{0.49\linewidth}
\centering
\includegraphics[width=\linewidth]{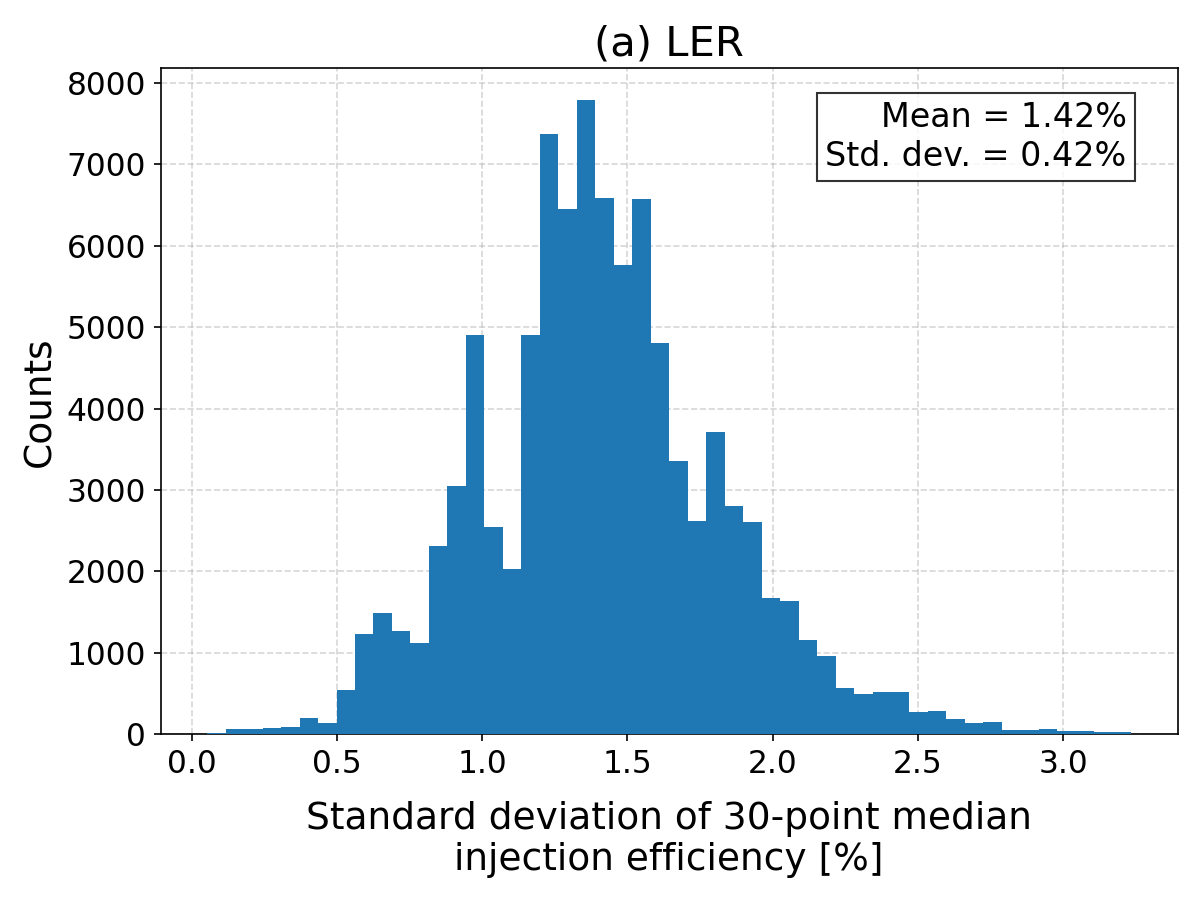}
\end{minipage}
\hfill
\begin{minipage}{0.49\linewidth}
\centering
\includegraphics[width=\linewidth]{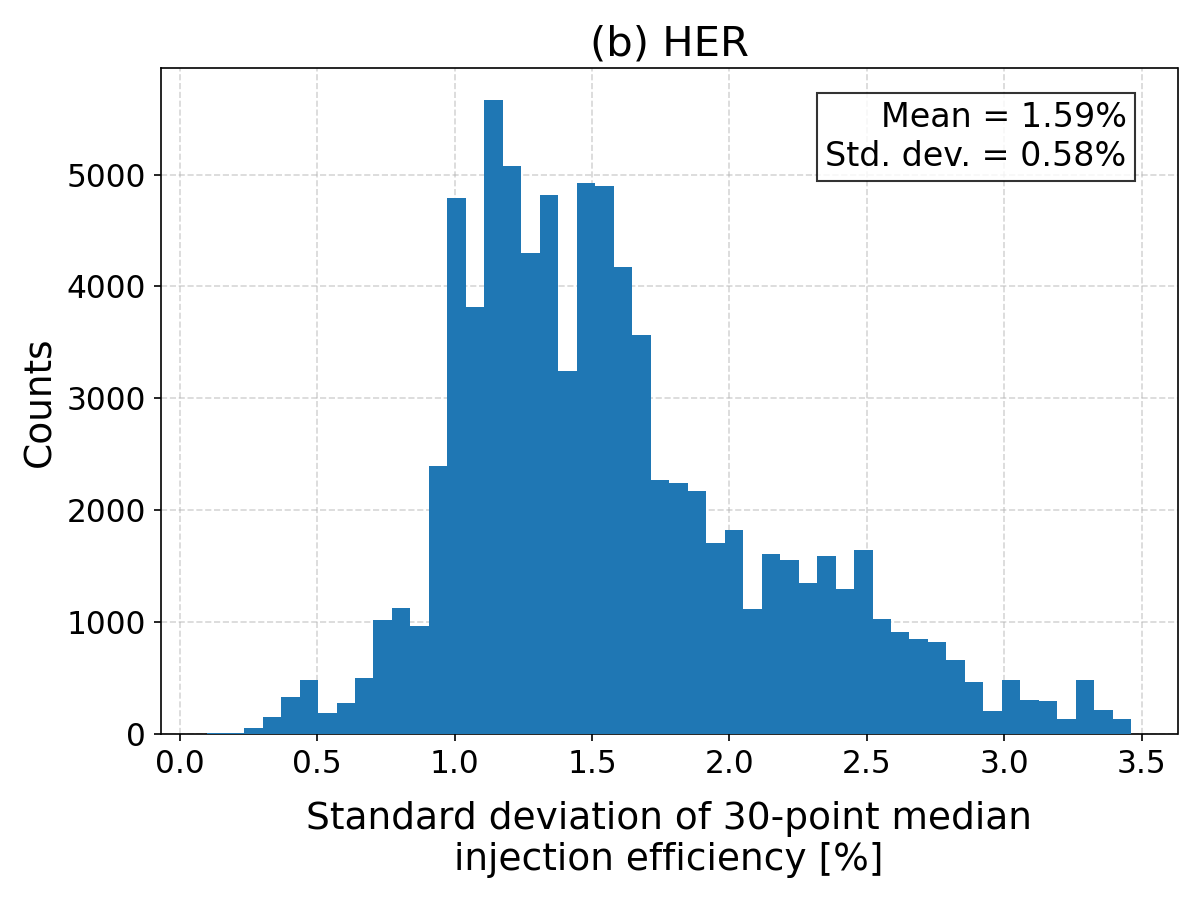}
\end{minipage}

\caption{
Distributions of the standard deviation of the 30-point median injection efficiency evaluated using 30-minute moving windows during stable beam operation: (a) LER and (b) HER.
}
\label{fig:median_hist}
\end{figure}

Since one Bayesian optimization run takes approximately 30 minutes, the statistical fluctuation of the injection efficiency was evaluated on the same time scale. For this purpose, periods of stable beam operation without tuning were selected from one week of operation from December 1 to December 7, 2024. The total selected periods were 41.2 hours for the LER and 34.5 hours for the HER. To evaluate the fluctuation on the time scale corresponding to one Bayesian optimization run, a 30-minute moving window was applied to the time series of the median injection efficiency during stable beam operation. For each window position, the standard deviation of the values contained in the window was calculated. Figure~\ref{fig:median_hist} shows the distribution of the obtained standard deviations, which represents the fluctuation of the injection efficiency on a 30-minute time scale. For the LER, the mean of this distribution was 1.41\%, with a standard deviation of 0.41\%. For the HER, the corresponding values were 1.6\% and 0.58\%, respectively. As a conservative estimate of the fluctuation during one optimization run, the mean plus three standard deviations of these distributions was used. This gives 2.6\% for the LER and 3.3\% for the HER. Furthermore, assuming that the injection efficiency itself may fluctuate by up to three times these values, the upper bounds of the fluctuation were estimated to be 7.9\% for the LER and 10.0\% for the HER. Therefore, in this study, improvements larger than these values were regarded as unlikely to be explained solely by statistical fluctuation and were considered statistically significant improvements attributable to Bayesian optimization.

\subsection{Verification of Reproducibility}
\label{Verification of Reproducibility}

In this section, we evaluate the reproducibility of the optimization results by conducting multiple optimization runs over a 2–3~h period, during which the beam conditions were assumed to remain constant. Each run was conducted using the same parameter domains and settings. Here, reproducibility refers to whether the parameter sets that improve the injection efficiency are consistently located within the same or nearby regions of the parameter space across different runs. The experiments were conducted three times for the HER and twice for the LER, with the HER results presented in this section.

The measurements for the HER were conducted on December 14, 2024 under beam collision conditions, with stored currents of 1080 mA in the HER and 1350 mA in the LER. The parameter domains were set as follows: $y$-position in $[0.9, 0.92]$ mm, $y$-angle in $[-0.025, -0.005]$ mrad, and Septum Angle in $[0.862, 0.882]$ mrad. The acquisition function used in the Bayesian optimization was Expected Improvement (EI).

\begin{figure}[htbp]
\centering
\includegraphics[width=\linewidth]{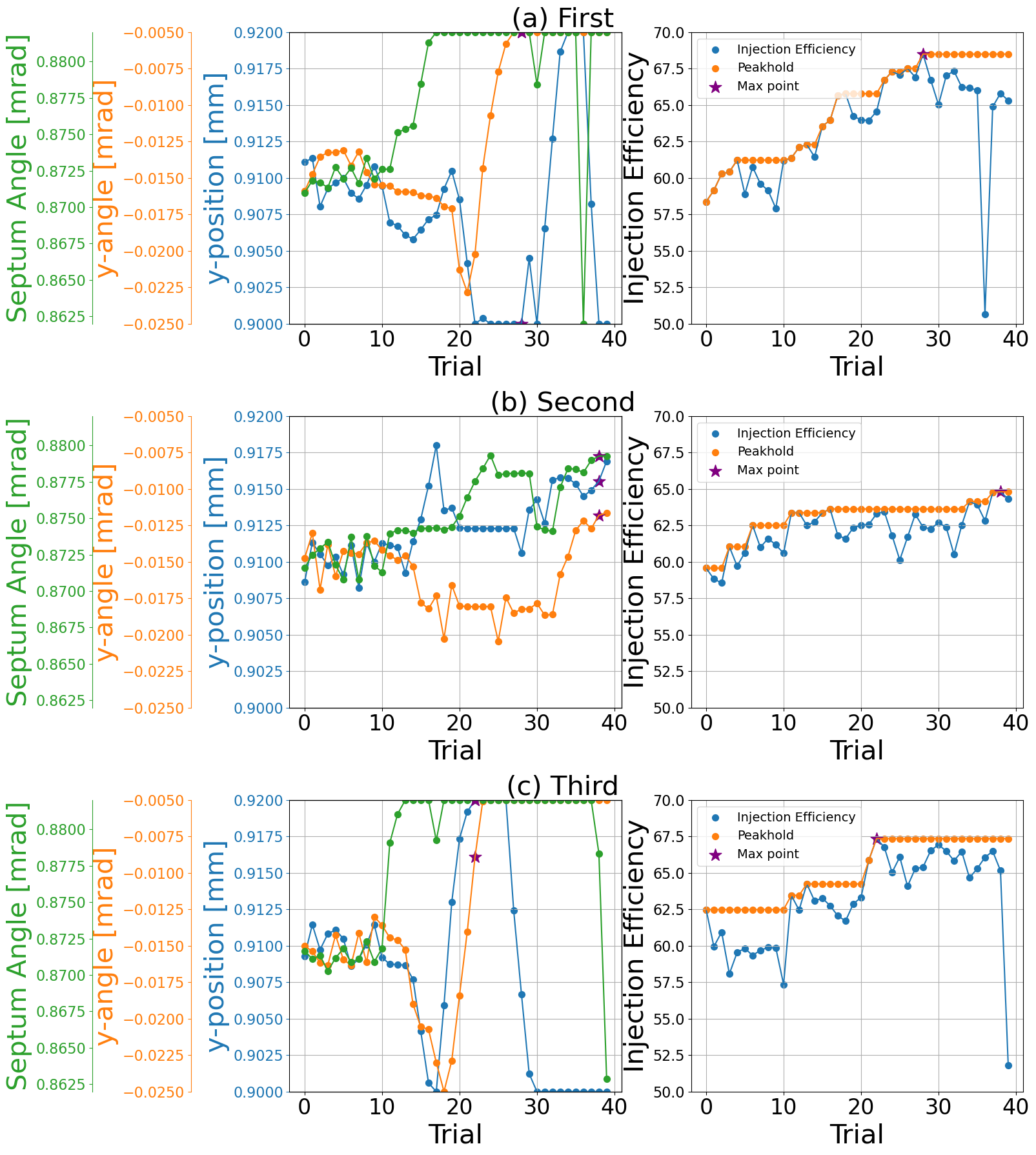}
\caption{Results of three optimization runs for the HER conducted under identical beam conditions and parameter domains.}
\label{fig:repro_compare}
\end{figure}

Figure~\ref{fig:repro_compare} shows the results of three optimization runs conducted under identical beam conditions and parameter domains for the HER. Panels (a)–(c) each consist of two subplots: the left panels display the variation in tuning parameters across trials, whereas the right panels show the injection efficiency for each trial. In the right panels, the blue line represents the injection efficiency, the orange line indicates the peakhold (the highest injection efficiency recorded up to that point), and the purple marker highlights the trial at which the maximum value was observed. The first ten points were generated based on the initialization conditions, and the subsequent 30 points were selected using Bayesian optimization. For convenience, these initial points are also counted in the total trial number. A comparison of the three runs revealed that the maximum injection efficiency was achieved when the $y$-angle and Septum Angle were greater than their initial values. In contrast, changes in $y$-position did not lead to significant variations in the injection efficiency.

\begin{figure}[htbp]
\centering
\includegraphics[width=\linewidth]{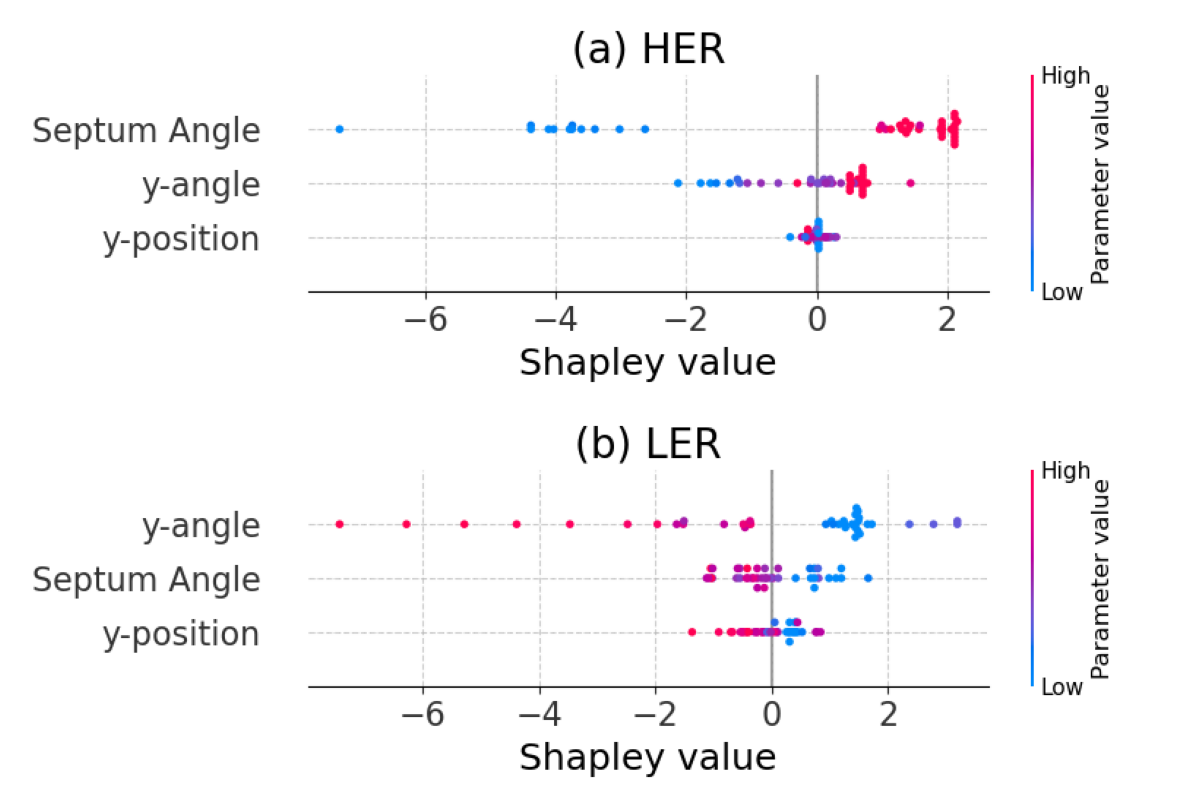}
\caption{Plots of tuning parameters and corresponding Shapley values in the optimization results for HER (a) and LER (b). }

\label{fig:SHAP_compare}
\end{figure}

To evaluate how the three injection-tuning parameters contribute to the injection efficiency, we performed an analysis using SHapley Additive exPlanations (SHAP)~\cite{lundberg2017unified}. SHAP is a method for interpreting the output of machine-learning models by quantitatively assessing the impact of each parameter on model prediction in terms of Shapley values. The Shapley values are interpreted as relative contributions within the selected control bounds, rather than as absolute machine sensitivities. As an example, Fig.~\ref{fig:SHAP_compare} shows the Shapley values based on the optimization results for both the HER and LER. In these plots, the colormap indicates the actual parameter values, and the Shapley values reflect the extent to which each parameter contributes to either an increase or decrease in the injection efficiency. In panel (a), it can be observed that higher values of the Septum Angle and y-angle led to increasing the injection efficiency, whereas lower values resulted in reduced efficiency. Conversely, panel (b) indicates that lower values of the Septum Angle and y-angle contribute to higher injection efficiency. In both cases,the y-position appeared to have little impact on the variation in the injection efficiency.

\begin{figure}[htbp]
\centering
\includegraphics[width=\linewidth]{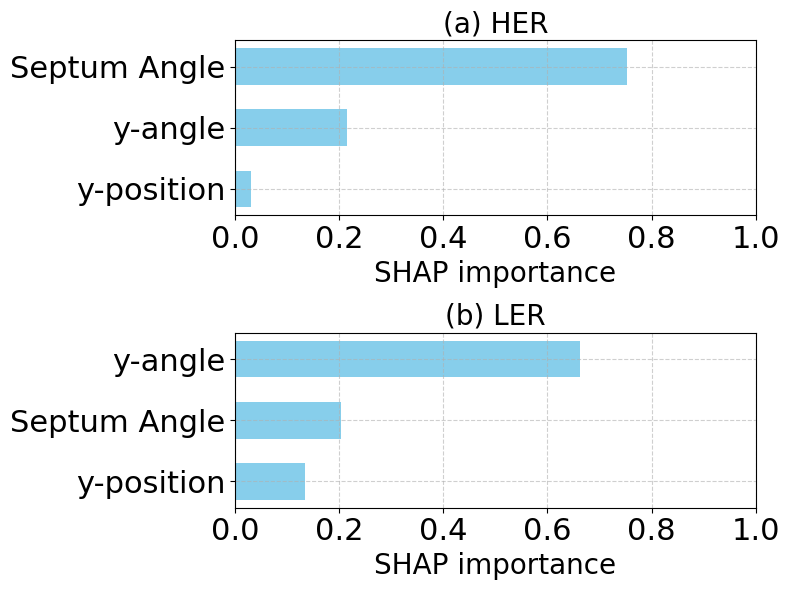}
\caption{SHAP importance for the optimization results of HER (a) and LER (b).}
\label{fig:SHAP_importance}
\end{figure}

Next, we conducted an analysis using SHAP importance. SHAP importance is a metric that quantifies the contribution of each parameter to a model's predictions by summing the absolute Shapley values for each parameter and normalizing the total to one. Figure~\ref{fig:SHAP_importance} shows the SHAP importance in the optimization of the HER (a) and LER (b). The results revealed that the contribution of the y-position to the injection efficiency was relatively small compared to those of the Septum Angle and y-angle. Moreover, the Septum Angle exhibited the highest importance in the HER, whereas the y-angle was the most influential parameter in the LER.

A similar analysis was conducted for all five optimization runs discussed in this section: three for the HER and two for the LER. In each case, the contribution of the y-position was consistently smaller than that of the Septum Angle and y-angle.

In conclusion, this section shows that parameters with high importance in injection tuning exhibit reproducibility; that is, the parameter settings that improve the injection efficiency consistently lie within the same or nearby regions across multiple optimization runs.

\subsection{Effect of Proximal Biasing Level}
\label{Effect of Proximal Biasing Level}

\begin{figure}[htbp]
\centering

\includegraphics[width=\linewidth]{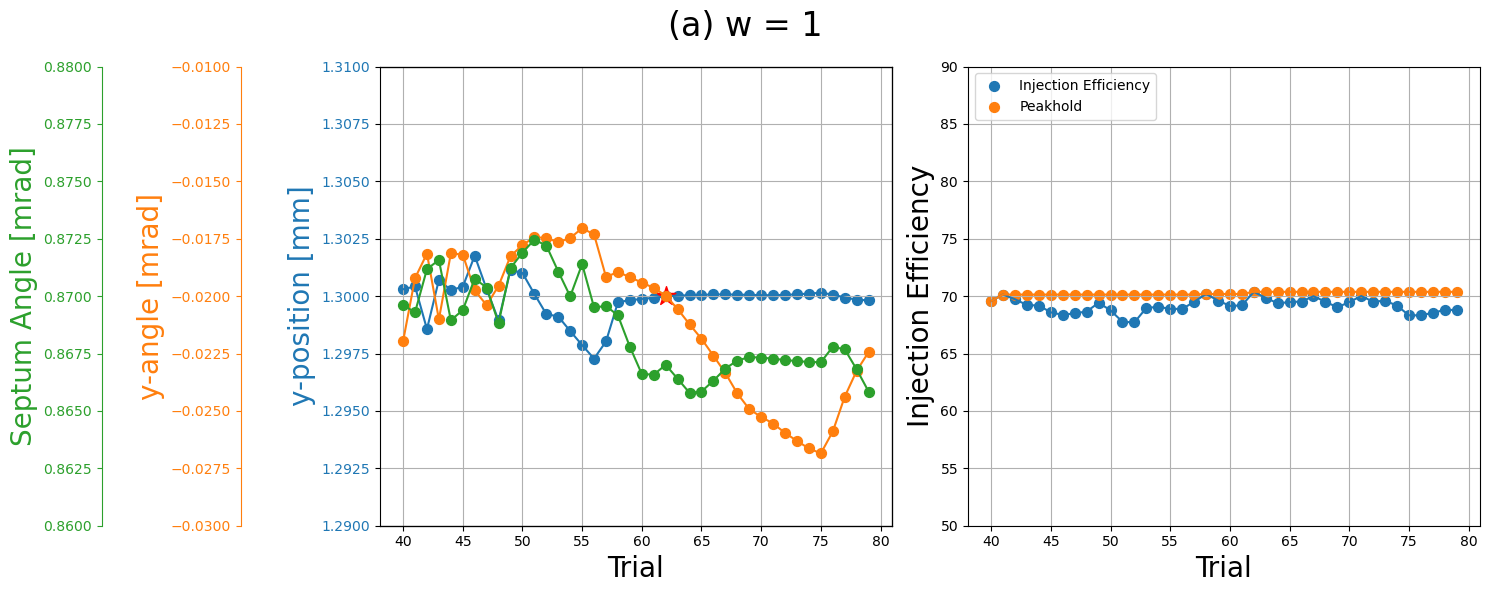}

\vspace{0.3cm}

\includegraphics[width=\linewidth]{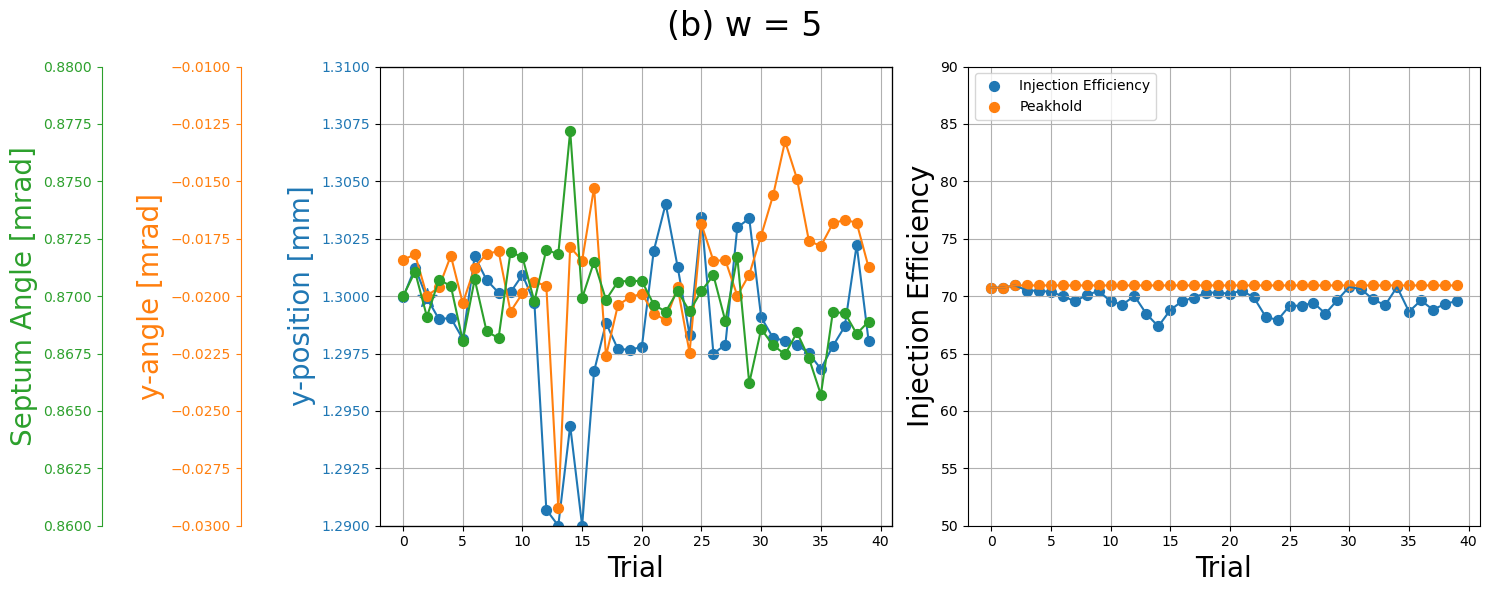}

\caption{Comparison of optimization results using UCB ($\beta = 2$): (a) proximal biasing factor set to 1 and (b) proximal biasing factor set to 5.}
\label{fig:proximal_compare}
\end{figure}

This section compares the effects of different proximal biasing factors on optimization speed. The acquisition function was UCB ($\beta = 2$), and we examined two settings: proximal biasing factors of 1 and 5. Figure~\ref{fig:proximal_compare} shows the results for each condition, with (a) corresponding to a factor of 1 and (b) corresponding to a factor of 5. Comparing the two plots, we observed that when the biasing factor was set to 1, the parameter changes between trials were smaller than when it was set to 5. In practice, the total optimization time was 26~min for a biasing factor of 1 and 24~min for a factor of 5, indicating that a higher biasing factor reduced the total time by approximately 2~min. No significant difference in optimization performance was observed between the two cases.

\begin{figure}[htbp]
\centering
\includegraphics[width=\linewidth]{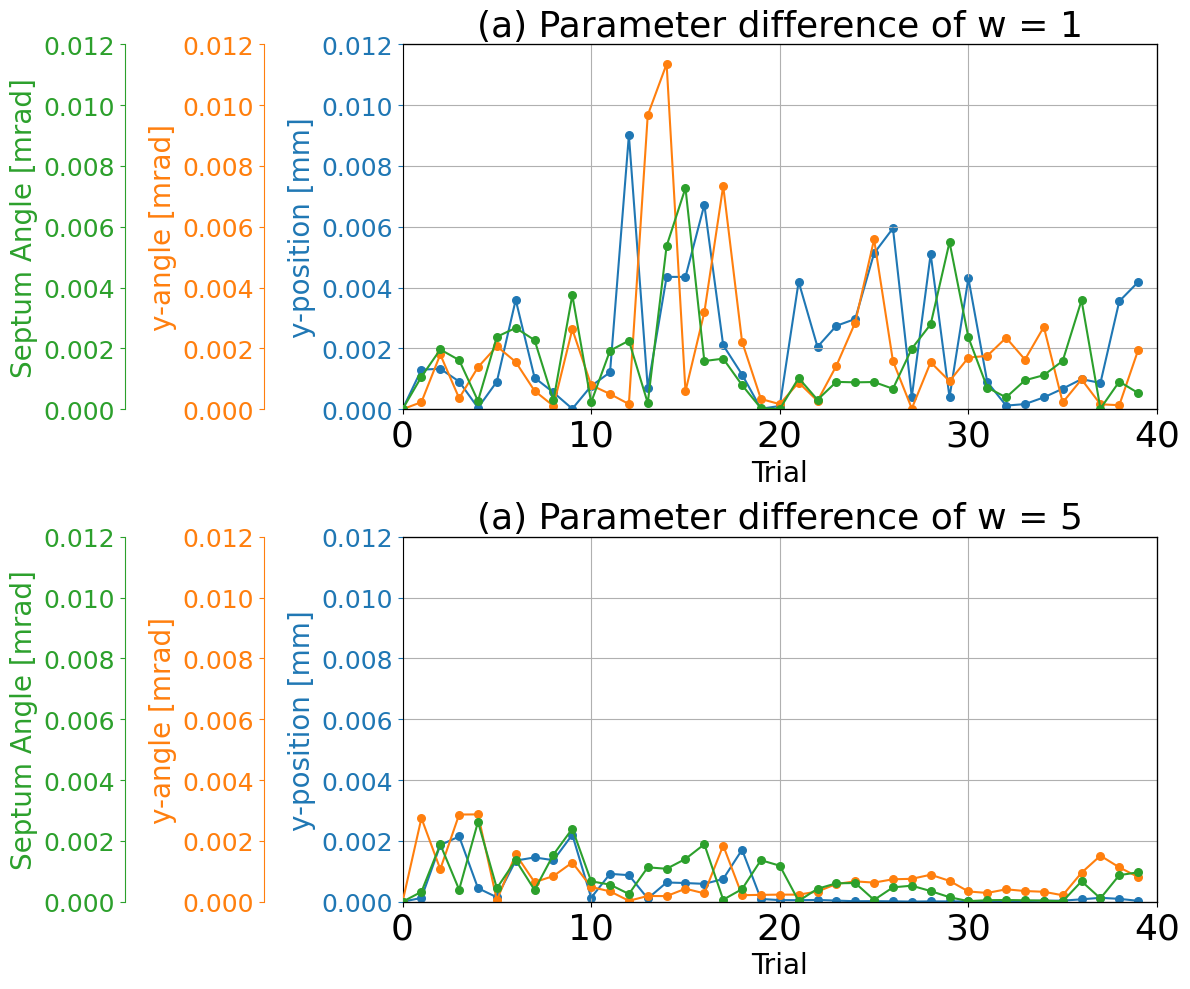}
\caption{Absolute per-trial changes in parameter values using the UCB acquisition function ($\beta = 2$), with the proximal biasing factor set to 1 (a) and 5 (b).}
\label{fig:difference_compare}
\end{figure}

Figure~\ref{fig:difference_compare} shows the amount of parameter changes across trials, plotted as the absolute difference in parameter values for each trial. When the proximal biasing factor was set to 1, the parameter changes were generally larger than those observed with a factor of 5. Because the parameter step size was limited to 0.0015 mm or mrad per second for each of the three parameters by the step-by-step variation, any change exceeding this value incurred a wait time proportional to the excess. For instance, a change of 0.012 requires approximately 8~s. Such delays accumulated over the course of 40 trials, resulting in a total optimization time more than 2~min longer than the case with a biasing factor of 5. A similar trend was observed when the acquisition function was set to the EI; comparing biasing factors of 0 and 5, the total optimization time differed by approximately 5~min.

Subsequently, we considered an appropriate setting for the proximal biasing factor in the optimization. This factor balances the trade-off between optimization performance and optimization time. When the proximal biasing factor was set to 5, the maximum parameter change observed in Fig.~\ref{fig:difference_compare} was 0.003 mrad, corresponding to a movement time of 2~s. This duration was not dominant compared to the approximately 30~s required to measure the injection efficiency 30 times, suggesting that increasing the biasing factor beyond 5 would be unlikely to yield significant benefits. Furthermore, no degradation in the optimization performance was observed with a setting of 5. Therefore, a proximal biasing factor of 5 was considered appropriate for the segmented optimization scheme used in this study.

\subsection{Test Deployment of the Injection-Tuning Tool}
\label{Test Deployment of the Injection Tuning Tool}

\begin{figure*}[htbp]
\centering
\includegraphics[width=0.8\linewidth]{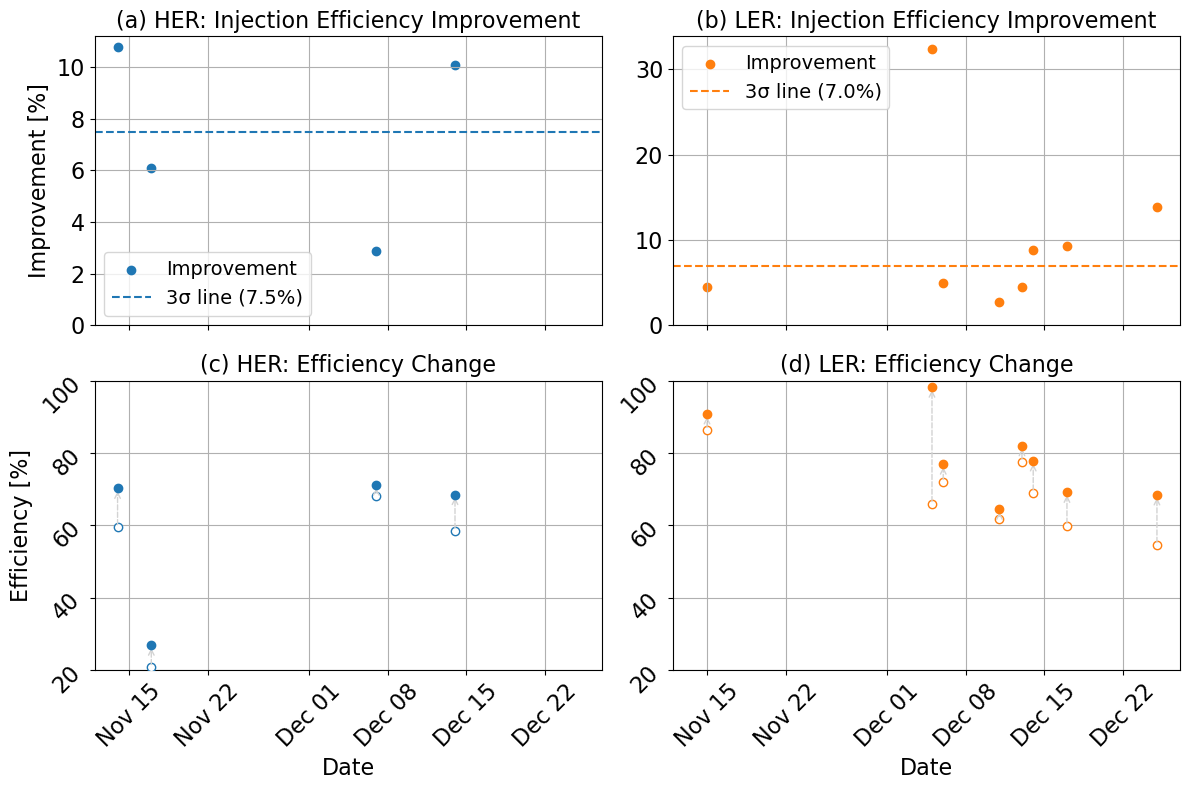}
\caption{Optimization results from the test deployment between November 14 and December 25, 2024. The circles represent the injection efficiency before tuning, while the filled circles represent the injection efficiency after tuning.}
\label{fig:improve_summery}
\end{figure*}

From November 14 to December 25, 2024, test deployments of the injection-tuning tool were conducted over four days for the HER and eight days for the LER. The results for each measurement day are presented in Fig.~\ref{fig:improve_summery}. Panels (a) and (b) show the improvements in the injection efficiency for the HER and LER, respectively. Panels (c) and (d) present the injection efficiencies before and after tuning, respectively, on the corresponding days.

Next, we discuss the optimization performance of the proposed tool. In Section~\ref{Uncertainty in Injection Efficiency}, a statistically significant improvement in the injection efficiency through Bayesian optimization was defined as an increase of at least 10.0\% for the HER and 7.9\% for the LER. Based on the results shown in Fig.~\ref{fig:improve_summery}, two out of four optimization runs for the HER and four out of eight runs for the LER achieved statistically significant improvements.

During SuperKEKB operation, injection is performed continuously throughout the day in top-up mode, and injection tuning is typically carried out more than five times per day. Therefore, a method that can safely improve the injection efficiency without introducing additional lost time is operationally valuable. In particular, the improvement achieved in the LER injection tuning on December 25 was operationally significant. The injection efficiency was improved from 54.5\% to 68.4\%, which enabled operation with an LER beam current of 1700~mA, the highest current ever achieved at SuperKEKB at that time. This contributed to the renewal of the world record for instantaneous luminosity for the first time in two and a half years.

Additionally, we evaluated the safety of the proposed tool with respect to beam aborts. During injection, if the beam-background level exceeds the threshold, beam loss monitors near the Belle~II detector issue an abort trigger. Throughout all runs conducted in this study, no beam aborts were triggered, indicating that the combination of segmented region optimization and the step-by-step process ensured a high level of operational safety.

\begin{figure}[htbp]
\centering
\includegraphics[width=\linewidth]{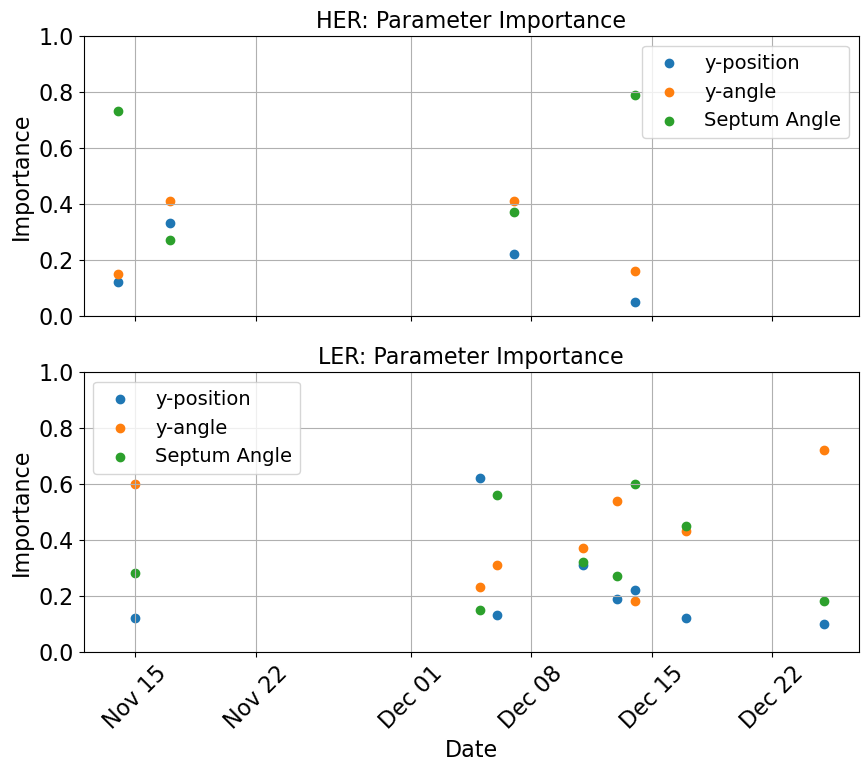}
\caption{SHAP importance during the test deployment conducted from November 14 to December 25, 2024.}
\label{fig:importance_summery}
\end{figure}

Figure~\ref{fig:importance_summery} shows the SHAP importance of each parameter for each measurement day during the test deployment of the injection-tuning tool. The figure reveals that the parameters contributing to the improvements in the injection efficiency varied from day to day. In particular, among the six days with statistically significant improvements, each of the three parameters (y-position, y-angle, and Septum Angle) appeared to be the most influential on at least one day, indicating that the dominant contributing parameter differed depending on the measurement day. Moreover, in both the HER and LER rings, the y-angle and Septum Angle contributed more frequently to efficiency improvements than the y-position. On the day with the smallest efficiency gain (December 11), the SHAP importance was nearly evenly distributed among the three parameters (approximately one-third each), suggesting that no single parameter played a dominant role in enhancing the injection efficiency on that day.

\section{Conclusion}
\label{Conclusion}

Achieving higher luminosity at the SuperKEKB accelerator requires further advancement of the injection-tuning system. In this study, we developed an injection-tuning tool based on Bayesian optimization.

To adapt Bayesian optimization for the injection tuning, three key features were implemented. First, the segmented region optimization avoids high-beam-loss parameter regions by initially restricting the search to a small domain and gradually expanding it toward directions that yield a higher injection efficiency. Second, the step-by-step variation prevent beam aborts by limiting the amount of change in tuning parameters. Third, the proximal biasing method reduces the overall optimization time. By performing the segmented region optimization and applying the step-by-step variation, no injection-related beam aborts occurred during any of the optimization runs conducted in November-December 2024. Furthermore, the use of the proximal biasing method successfully reduced the optimization time by 2–5~min. 

The tool was tested over four days for the HER and eight days for the LER. Statistically significant improvements in the injection efficiency, exceeding the level of statistical fluctuation, were observed on two days for the HER and four days in the LER. Additionally, an analysis of parameter importance using SHAP demonstrated that the contribution of the tuning parameters to the injection efficiency varied significantly from day to day.

Although only three parameters were optimized in this first study, these parameters are among the most sensitive knobs for injection tuning, where inappropriate settings or rapid changes can cause beam loss, beam aborts, or even equipment damage. Therefore, the significance of this work lies in demonstrating the safe and reliable optimization of such sensitive parameters under machine-protection constraints, while physics data taking was ongoing and without contaminating the physics data. To the best of our knowledge, this is the first successful application of Bayesian optimization to injection tuning in an operational top-up collider, and this demonstration provides a useful benchmark for future top-up electron--positron colliders, such as FCC-ee and CEPC.

Since the SuperKEKB accelerator exhibits machine drift during operation, an important direction for future work is to incorporate such time-dependent machine conditions into the Bayesian optimization framework.

\section*{Acknowledgment}

The authors would like to express their sincere gratitude to all those involved in this research, including the members of the KEK accelerator facility, accelerator operators, and the Belle~II collaboration. In particular, we are deeply grateful to Prof. Naoko Iida and Prof. Masanori Sato of the KEK accelerator group for their valuable guidance regarding the injection system and injection tuning, and to Associate Prof. Takuya Natsui for his insightful advice on the implementation of Bayesian optimization.

S.~Kato and G.~Mitsuka work was supported by Japan Society for the Promotion of Science (JSPS) International Leading Research (Grant Number: JP22K21347), and JSPS Core-to-Core Program (Grant Number: JPJSCCA20230004). In addition, S.~Kato work was supported by The University of Tokyo Advanced AI Talent Development to Lead the Next-Generation Intelligent Society (BOOST NAIS) (Grant Number: JPMJBS2418).

\bibliographystyle{unsrt}
\bibliography{references}

@misc{SuperKEKB:TDR_overview,
  author  = {SuperKEKB Collaboration},
  title   = {SuperKEKB Design Report, Overview},
  year    = {2019},
  note    = {Available at: \url{https://www-linac.kek.jp/linac-com/report/skb-tdr/2_MachineParameters.pdf}}
}

@misc{SuperKEKB:TDR_BT,
  author  = {SuperKEKB Collaboration},
  title   = {SuperKEKB Design Report, Beam Transport},
  year    = {2020},
  note    = {Available at: \url{https://www-linac.kek.jp/linac-com/report/skb-tdr/10_BT_Revised2020_10_20.pdf}}
}

@misc{SuperKEKB:TDR_monitor,
  author  = {SuperKEKB Collaboration},
  title   = {SuperKEKB Design Report, Beam Instrumentation},
  year    = {2020},
  note    = {Available at: \url{https://kds.kek.jp/event/15914/contributions/28491/attachments/136900/166652/BeamInstrumentation2020_10_19s.pdf}}
}

@inproceedings{NatsuiLCWS2024,
  author       = {Natsui, T.},
  title        = {{SuperKEKB positron beam tuning using machine learning}},
  booktitle    = {EPJ Web of Conferences},
  volume       = {315},
  pages        = {02004},
  year         = {2024},
  doi          = {10.1051/epjconf/202431502004},
  url          = {https://www.epj-conferences.org/articles/epjconf/abs/2024/25/epjconf_lcws2024_02004/epjconf_lcws2024_02004.html},
  note         = {International Workshop on Future Linear Colliders (LCWS2024), Tokyo, Japan, July 8–11, 2024},
}

@misc{proximal,
    author  = {Roussel, R. and Edelen, A.},
    title   = {Proximal Biasing for Bayesian Optimization and Characterization of Physical Systems},
    year    = {2021},
    url     = {https://ml4physicalsciences.github.io/2021/files/NeurIPS_ML4PS_2021_82.pdf},
    note    = {NeurIPS Workshop on Machine Learning and the Physical Sciences},
}

@book{rasmussen2006gaussian,
  title={Gaussian Processes for Machine Learning},
  author={Rasmussen, C. E. and Williams, C. K. I.},
  year={2006},
  publisher={MIT Press},
  address={Cambridge, MA},
  isbn={9780262182539},
  url={http://www.gaussianprocess.org/gpml/}
}

@article{10.1093/comjnl/7.4.308,
    author = {Nelder, J. A. and Mead, R.},
    title = {A Simplex Method for Function Minimization},
    journal = {The Computer Journal},
    volume = {7},
    number = {4},
    pages = {308-313},
    year = {1965},
    month = {01},
    issn = {0010-4620},
    doi = {10.1093/comjnl/7.4.308},
    url = {https://doi.org/10.1093/comjnl/7.4.308},
    eprint = {https://academic.oup.com/comjnl/article-pdf/7/4/308/1013182/7-4-308.pdf},
}

@article{PhysRevAccelBeams.27.084601,
  title = {Machine-learning approach for operating electron beam at KEK electron/positron injector linac},
  author = {Mitsuka, G. and Kato, S. and Iida, N. and Natsui, T. and Satoh, M.},
  journal = {Phys. Rev. Accel. Beams},
  volume = {27},
  issue = {8},
  pages = {084601},
  numpages = {12},
  year = {2024},
  month = {Aug},
  publisher = {American Physical Society},
  doi = {10.1103/PhysRevAccelBeams.27.084601},
  url = {https://link.aps.org/doi/10.1103/PhysRevAccelBeams.27.084601}
}

@incollection{Mockus1978,
  author    = {Mockus, J. and Tiesis, V. and Zilinskas, A.},
  title     = {The application of Bayesian methods for seeking the extremum},
  booktitle = {Towards Global Optimization},
  volume    = {2},
  pages     = {117--129},
  year      = {1978},
  publisher = {Elsevier},
  editor    = {L. C. W. Dixon and G. P. Szego}
}

@book{Mockus1989,
  author    = {Mockus, J.},
  title     = {Bayesian Approach to Global Optimization: Theory and Applications},
  year      = {1989},
  publisher = {Kluwer Academic Publishers},
  address   = {Dordrecht, Netherlands},
  isbn      = {978-94-015-7793-1},
  doi       = {10.1007/978-94-015-7793-1}
}

@article{Roussel:2023yin,
    author = "Roussel, R. and others",
    title = "{Bayesian optimization algorithms for accelerator physics}",
    eprint = "2312.05667",
    archivePrefix = "arXiv",
    primaryClass = "physics.acc-ph",
    reportNumber = "FERMILAB-PUB-23-0846-AD",
    doi = "10.1103/PhysRevAccelBeams.27.084801",
    journal = "Phys. Rev. Accel. Beams",
    volume = "27",
    number = "8",
    pages = "084801",
    year = "2024"
}

@article{striplineBPM,
author = {Suwada, T. and Kamikubota, N. and Fukuma, H. and Akasaka, N. and Kobayashi, H.},
year = {2000},
month = {02},
pages = {307--319},
title = {Stripline-type beam-position-monitor system for single-bunch electron/positron beams},
volume = {440},
journal = {Nuclear Instruments and Methods in Physics Research Section A: Accelerators, Spectrometers, Detectors and Associated Equipment},
doi = {10.1016/S0168-9002(99)00960-2}
}

@inproceedings{NIPS2011_86e8f7ab,
 author = {Bergstra, J. and Bardenet, R. and Bengio, Y. and K\'{e}gl, B.},
 booktitle = {Advances in Neural Information Processing Systems},
 editor = {J. Shawe-Taylor and R. Zemel and P. Bartlett and F. Pereira and K.Q. Weinberger},
 pages = {},
 publisher = {Curran Associates, Inc.},
 title = {Algorithms for Hyper-Parameter Optimization},
 url = {https://proceedings.neurips.cc/paper_files/paper/2011/file/86e8f7ab32cfd12577bc2619bc635690-Paper.pdf},
 volume = {24},
 year = {2011}
}

@misc{hansen2023cmaevolutionstrategytutorial,
      title={The CMA Evolution Strategy: A Tutorial}, 
      author={Hansen, N.},
      year={2023},
      eprint={1604.00772},
      archivePrefix={arXiv},
      primaryClass={cs.LG},
      url={https://arxiv.org/abs/1604.00772}, 
}

@inproceedings{lundberg2017unified,
  title={A Unified Approach to Interpreting Model Predictions},
  author={Lundberg, S. M. and Lee, S.-I.},
  booktitle={Advances in Neural Information Processing Systems},
  volume={30},
  year={2017}
}

@inproceedings{eriksson2019scalable,
  title     = {Scalable Global Optimization via Local Bayesian Optimization},
  author    = {Eriksson, David and Pearce, Michael and Gardner, Jacob and Turner, Ryan D. and Poloczek, Matthias},
  booktitle = {Advances in Neural Information Processing Systems},
  volume    = {32},
  year      = {2019}
}

\end{document}